\documentclass[aps,pra,groupedaddress,twocolumn,showpacs,longbibliography]{revtex4-1}
\usepackage{amsmath}
\usepackage{amsthm, amssymb}
\usepackage{bbold}
\usepackage{graphicx}
\usepackage{color}
\usepackage{times, verbatim}
\usepackage{tikz}
\usepackage{hyperref}
\hypersetup{
       colorlinks=true,
}

\begin{document}

\title{Degeneracy Implies Non-abelian Statistics}

\author{Eric C. Rowell$^{1}$ and Zhenghan Wang$^{2}$}
\affiliation{$^{1}$Department of Mathematics, Texas A\&M University, College Station, TX 77843-3368, USA}
\affiliation{$^{2}$Microsoft Station Q and Department of Mathematics, University of California, Santa Barbara, California 93106, USA}

\begin{abstract}
A non-abelian anyon can only occur in the presence of ground state degeneracy in the plane.  It is conceivable that for some strange anyon with quantum dimension $>1$ that the resulting representations of all $n$-strand braid groups $B_n$ are overall phases, even though the ground state manifolds for $n$ such anyons in the plane are in general Hilbert spaces of dimensions $>1$.  We observe that degeneracy is all that is needed: for an anyon with quantum dimension $>1$ the non-abelian statistics cannot all be overall phases on the degeneracy ground state manifold. Therefore, degeneracy implies non-abelian statistics, which justifies defining a non-abelian anyon as one with quantum dimension $>1$.  Since non-abelian statistics presumes degeneracy, degeneracy is more fundamental than non-abelian statistics.

\end{abstract}

\date{\today}


\maketitle

\section{Introduction}

In 1991, the potential realization of non-abelian statistics in fractional quantum Hall states was proposed \cite{mr91,wen91}.  Recently, the Majorana zero-mode version of non-abelian statistics has been intensively pursued in experiments using nanowires (see Ref. \cite{dfn15}).  More generally, non-abelian statistics occur in topological phases matter---quantum phases of matter that exhibit topological orders \footnote{Our terminologies are as follows: a topological phase of matter is an equivalence class of Hamiltonians or states without any phase transitions among them, while a topological order is the topological quantum field theory or unitary modular tensor category encoded in the Hamiltonian or state class at low energy}.  A direct observation of non-abelian statistics will be to braid the non-abelian objects.  But an easier experiment than braiding non-abelian objects is to observe the Ising fusion rule $\sigma\otimes \sigma = 1\oplus \psi$.  This Ising fusion rule implies degeneracy, and more generally any anyon with quantum dimension $d>1$ has degeneracy.  Does non-
abelian statistics follow from degeneracy? \footnote{The second author thanks Jason Alicea for asking him this question in June, 2015.} In this letter, we point out that indeed degeneracy implies non-abelian statistics.  Since degeneracy is a prerequisite for non-abelian statistics, degeneracy is more fundamental than non-abelian statistics in a sense.  Without our observation, replacing non-abelian statistics by degeneracy is unjustified.

Non-abelian statistics is a fundamentally new form of particle interactions.  This \lq\lq spooky action" is a manifestation of entanglement in the degenerate ground states---{\it the} characteristic attribute of quantum mechanics according to Sch\"odinger.  Besides its general interest as a new form of particle interaction, non-abelian statistics underlies the idea of topological quantum computation---the braiding matrices are inherently fault-tolerant quantum circuits \cite{kitaev97, freedman98, fklw03}.  Therefore, it is crucial to confirm non-abelian statistics by experiments.  The Ising fusion rule $\sigma\otimes \sigma = 1\oplus \psi$ is amenable to experimental test in nanowire technology now \cite{jason15}.  Our result means that if we can verify a non-abelian fusion rule by experiments, then on one hand, it is also a verification of non-abelian particle interactions, and on the other hand, it establishes the feasibility of the construction of a topological quantum computer.

\section{Anyon Models}

Anyons are topological quantum fields materialized as finite energy particle-like excitations in topological phases of matter.  Like particles, they can be moved, but cannot be created or destroyed by local operators alone.  Two anyons have the same anyon type or topological charge if they differ by local operators.  There are two equivalent ways to model anyon systems.  We can focus on the ground state manfold $V(Y)$ of an anyonic system on any possible space $Y$, and then the anyon system is modeled in low energy by a $(2+1)$-dimensional topological quantum field theory (TQFT)  $\{V(Y)\}$.  An alternative is to consider the fusion and braiding structures of all elementary excitations in the plane.  The anyon system is then equivalently modeled by a unitary modular (tensor) category $\mathcal{C}$.  The two notions $(2+1)$-TQFT and modular category are essentially the same \cite{tu94}.  Therefore, anyon systems can be modelled either by TQFTs or unitary modular categories.  In this letter, we will use
unitary modular categories to model anyon systems (see Ref. \cite{wa10}).

In the modular category model, an anyon $X$ is a simple object that abstracts an irreducible representation of some symmetry algebra.  The topological charge or anyon type $x$ of an anyon $X$ is an equivalent class of anyons \footnote{In general, it is important to distinguish anyon types with anyon objects.  We will use upper case letters to denote anyon objects and the corresponding lower cases to denote their types.}.  All possible topological charges in an anyon system form a finite label set $L=\{a,b,...\}$ with fusion rules $\{N_{ab}^c\}$, where $N_{ab}^c$ are non-negative integers \cite{wa10}.  The fusion rule $N_{ab}^c$ encodes the possible topological charges $c$ that will appear when two anyons of types $a,b$ are fused: if $N_{ab}^c=0$, then anyons of type $c$ will not appear; otherwise $N_{ab}^c>0$ and there are $N_{ab}^c$ different fusion channels for anyons $A,B$ to fuse to anyon $C$.  There is always a label $1$ in $L$ that corresponds to the ground state or
vacuum.  In the famous Ising theory, the label set is $L=\{1,\sigma,\psi\}$.  Usually, we write the fusion rules as a tensor-sum $a\otimes b = \oplus_{c\in L} N_{ab}^c c$.  There are always the trivial fusion rules $1\otimes x=x\otimes 1=x$.  In this tensor-sum notation, the non-trivial fusion rules for the Ising theory are $\sigma\otimes \sigma = 1\oplus \psi, \sigma\otimes \psi=\psi\otimes \sigma=\sigma, \psi\otimes \psi=1$.  The anyon $\sigma$ is called the Ising anyon.  The anyon $\psi$ is a fermion.  Generally, an anyon is self-dual if it has the same anyon type as its anti-particle.  Both $\sigma$ and $\psi$ are self-dual, hence $\psi$ is Majorana---a real fermion.

\section{Quantum Dimension and Degeneracy}

An important quantum number of an anyon $X$ is its quantum dimension $d_X$---a positive real number $\geq 1$.  The quantum dimension of an anyon can be easily computed from its fusion rules: regard all anyon types as unknown variables and the fusion rules as polynomial equations, then the maximal real solutions of these polynomial equations are the quantum dimensions.  For the Ising fusion rules, the quantum dimension of the Ising anyon $\sigma$ is $d_\sigma=\sqrt{2}$ and the Majorana fermion $d_\psi=1$.  The quantum dimension $d_X$ of an anyon $X$ determines the asymptotic growth rate when $n$ identical anyons $X$ are confined to the sphere: the dimension of the degeneracy ground state manifold $V_{X,n,1}$ grows as $d_X^n$ as $n\rightarrow \infty$.  Therefore, an anyon $X$ leads to degeneracy in the plane if and only if its quantum dimension $d_X >1$.

\section{Braid Groups and Non-abelian Statistics}

In two spatial dimensions, statistics of quasi-particles can be more general than bosons and fermions (see Ref. \cite{nayaktqc08}).  An exotic form of statistics is not an overall phase, but a unitary matrix: the overall change when two anyons $X$ in $n$ identical anyons $X$ are exchanged is a unitary matrix on the degeneracy ground state manifolds $V_{x,n,a}$ for some total topological charge $a$.  It follows that non-abelian statistics presumes degeneracy.

The ground state manifold $V_{x,n,a}$ is a representation of the $n$-stand braid group $B_n$. It is well-known that the braid group $B_n$ is generated by $n$ elementary braids $\{\sigma_i\}, i=1,2,...,n-1$.  {\bf Non-abelian statistics} means that the image of the representation for some $B_n$ is a non-abelian subgroup of the unitary group $U(V_{x,n,a})$.  It is conceivable that for some particular anyon $X$ with quantum dimension $>1$ that all representations of $B_n$ are overall phases, even though $V_{X,n,a}$ are Hilbert spaces of dimensions $>1$ for general $n$.  We will see below that this cannot occur.  At the end of Sec. IIB in \cite{bbcw14}, a weaker version of degeneracy implies non-abelan statistics is proved \footnote{There, it is proved that for any anyon $X$ with $d_X>1$, there is another anyon $Y$, possible $Y=X$, that the double braiding between $X$ and $Y$ is not a phase.  In particular, if an anyon system has only one type of non-abelian anyons, then its double braiding is not a phase.}.

\section{Degeneracy Implies Non-abelain Statistics}

When $n$ anyons $X$ are pinned in the plane, the ground state manifold $V_{X,n,a}$ for some total charge $a$ consists of exponentially closed degenerate ground states.  An orthonormal basis of $V_{X,n,a}$ is usually represented by labelled fusion trees (see Ref. \cite{wa10}).  The statistics of the anyon $X$ is computed by stacking braids on top of any state in $V_{X,n,a}$.  Physical states have to satisfy fusion rules at each trivalent vertex.

An anyon $X$ with $d_X>1$ is self-dual if and only if $X\otimes X=1\oplus Y\oplus...$ for some non-trivial anyon $Y$, which could have multiplicity.  Otherwise, there is a different anyon $X^*$ such that $X\otimes X^{*}=1\oplus Y\oplus...$ for some nontrivial anyon $Y$.

{\bf Theorem}: Suppose $X$ is an anyon with quantum dimension $d_X>1$. Then

1) If $X$ is self-dual, the image of the afforded representation of the $3$-strand braid group $B_3$ is non-abelian;

2) If $X$ is non-self-dual, the image of the afforded representation of the $4$-strand pure braid group ${P}_4$ is not trivial up to scalars.

If an anyon $X$ with $d_X>1$ is self-dual, to show that the $3$-strand braid group $B_3$ has an image which is non-abelian, we consider the following braid $b=\sigma^{-1}_2 \sigma^{-1}_1\sigma_2 \sigma_1$ in the $3$-strand braid group $B_3$.  Note the braid is the commutator of the two elementary braids $\sigma_1$ and $\sigma_2$.  We choose the representation $V_{X,3,X}$ \footnote{For convenience, we choose the total charge to be $X$.  The argument applies to other admissible total charges too.}.  Note that $\textrm{dim}V_{X,3,X}\geq 2$.  Starting with the state of three anyons $X$ in $V_{X,3,X}$ represented by the fusion tree below the bottom horizontal line, we braid the three anyons $X$ through the braid $b$.  We want to compute the resulting state after braiding $b$ with the constraint that the total charge of the first two anyons $X$ is $Y$  (See Fig. 1).

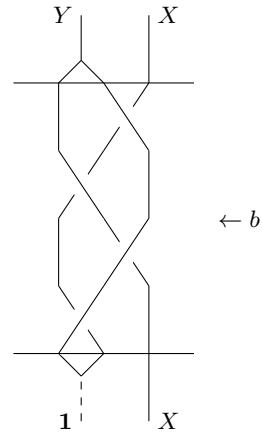
\begin{figure}[th!]
\begin{center}
\begin{tikzpicture}[scale=0.3]
\begin{scope}[yshift=0cm]
 \draw [dashed] (0,-2)--(0,0);
 \draw (0,0)--(1,1);
 \draw (0,0)--(-1,1);
 \draw (3, -2)--(3,1);
 \draw (0,-2) node[anchor=east] {$\mathbf{1}$};
 \draw (3,-2) node[anchor=west] {${X}$};
\end{scope}
\begin{scope}[yshift=3cm]
 \draw (1,-2)--(-1,1);
 \draw [white, line width=3mm] (-1,-2)--(1,1);
 \draw (-1,-2)--(1,1);
 \draw (3, -2)--(3,1);
 \draw (5, -2)--(-3,-2);
\end{scope}
\begin{scope}[yshift=6cm]
 \draw (3,-2)--(1,1);
 \draw [white, line width=3mm] (1,-2)--(3,1);
 \draw (1,-2)--(3,1);
 \draw (-1, -2)--(-1,1);
\end{scope}
\begin{scope}[yshift=9cm]
 \draw (-1,-2)--(1,1);
 \draw [white, line width=3mm] (1,-2)--(-1,1);
 \draw (1,-2)--(-1,1);
 \draw (3, -2)--(3,1);
\end{scope}
\begin{scope}[yshift=12cm]
 \draw (1,-2)--(3,1);
 \draw [white, line width=3mm] (3,-2)--(1,1);
 \draw (3,-2)--(1,1);
 \draw (-1, -2)--(-1,1);
 \draw (5, 1)--(-3,1);
\end{scope}
\begin{scope}[yshift=14cm]
 \draw (0,0)--(0,2);
 \draw (-1,-1)--(0,0);
 \draw (1,-1)--(0,0);
 \draw (3, -1)--(3,2);
 \draw (0,2) node[anchor=east] {${Y}$};
 \draw (3,2) node[anchor=west] {${X}$};
\end{scope}
\begin{scope}[xshift=7cm,yshift=7cm]
 \draw (0,0) node {$\leftarrow{b}$};
\end{scope}
\end{tikzpicture}

  \caption{(Color online) The self-dual case}
\label{fig:b}
\end{center}
\end{figure}

By sliding and twisting, we can deform the thick braided arc in Fig. 2 to the thick interval in the trivalent vertex without deforming the other arc, hence deforming the braided fusion tree to a trivalent vertex.  Therefore, up to an overall non-zero scalar $\alpha=\theta_X^2 R^{Y}_{XX}$, the resulting state is a trivalent vertex state (See Fig. 2).  If $Y$ has multiplicity, we choose a trivalent vertex state where the braiding $R^Y_{XX}$ acts as a scalar.

\begin{figure}[th!]
\begin{center}
\begin{tikzpicture}[scale=0.3]
  \begin{scope}
\begin{scope}[yshift=0cm]
 \draw [dashed] (0,-2)--(0,0);
 \draw [thick] (0,0)--(1,1);
 \draw [thick] (0,0)--(-1,1);
 \draw (3, -2)--(3,1);
 \draw (0,-2) node[anchor=east] {${1}$};
 \draw (3,-2) node[anchor=west] {${X}$};
\end{scope}
\begin{scope}[yshift=3cm]
 \draw [thick] (1,-2)--(-1,1);
 \draw [white, line width=3mm] (-1,-2)--(1,1);
 \draw [thick](-1,-2)--(1,1);
 \draw (3, -2)--(3,1);
 \draw (5, -2)--(-3,-2);
\end{scope}
\begin{scope}[yshift=6cm]
 \draw  (3,-2)--(1,1);
 \draw [white, line width=3mm] (1,-2)--(3,1);
 \draw [thick] (1,-2)--(3,1);
 \draw [thick] (-1, -2)--(-1,1);
\end{scope}
\begin{scope}[yshift=9cm]
 \draw [thick] (-1,-2)--(1,1);
 \draw [white, line width=3mm] (1,-2)--(-1,1);
 \draw (1,-2)--(-1,1);
 \draw [thick] (3, -2)--(3,1);
\end{scope}
\begin{scope}[yshift=12cm]
 \draw [thick] (1,-2)--(3,1);
 \draw [white, line width=3mm] (3,-2)--(1,1);
 \draw [thick](3,-2)--(1,1);
 \draw  (-1, -2)--(-1,1);
 \draw  (5, 1)--(-3,1);
\end{scope}
\begin{scope}[yshift=14cm]
 \draw (0,0)--(0,2);
 \draw (-1,-1)--(0,0);
 \draw [thick] (1,-1)--(0,0);
 \draw [thick] (3, -1)--(3,2);
 \draw (0,2) node[anchor=east] {${Y}$};
 \draw (3,2) node[anchor=west] {${X}$};
\end{scope}
 \end{scope}
\begin{scope}[xshift=8cm,yshift=7cm]
 \draw (0,0) node {$=~~\alpha$};
\end{scope}
\begin{scope}[xshift=13cm,yshift=9cm]
 \draw (0,-4)--(0,0);
 \draw [thick] (0,0)--(2,2);
 \draw (0,0)--(-2,2);
 \draw (0,-4) node[anchor=north] {${X}$};
 \draw (-2,2) node[anchor=east] {${Y}$};
 \draw (2,2) node[anchor=west] {${X}$};
\end{scope}
\begin{scope}[xshift=19cm,yshift=7cm]
 \draw (0,0) node {$\neq~~0$};
\end{scope}
\end{tikzpicture}
\label{fig:non-zero state}
\caption{Self-dual case: a non-zero state}
\end{center}
\end{figure}
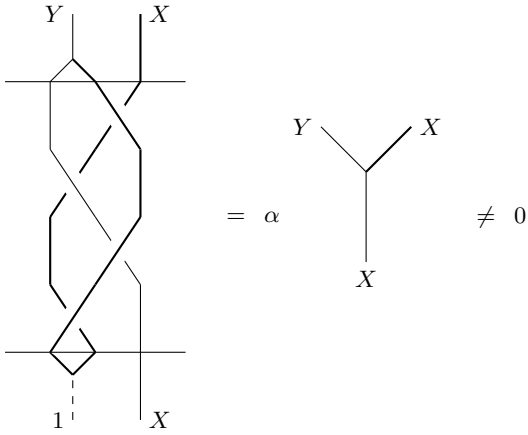

The trivalent vertex state is non-zero because the fusion rule is admissible.  On the other hand, if the braid $b$ has the same image as the identity up to some scalar $\gamma$, we can replace the braid $b$ between the two horizontal lines in Fig 1. by the identity braid.  Then the resulting state will be $0$ due to the no-tadpole rule (See Fig. 3).

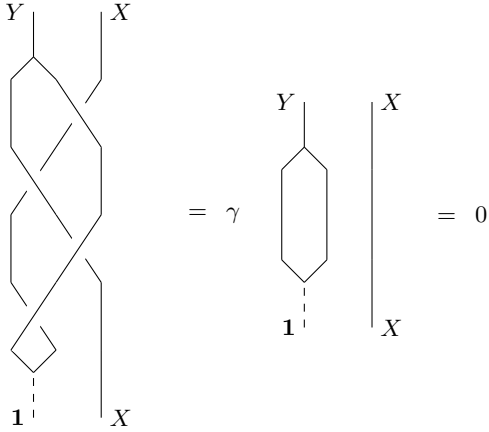
\begin{figure}[th!]
\begin{center}
\begin{tikzpicture}[scale=0.3]
 \begin{scope}
\begin{scope}[yshift=0cm]
 \draw [dashed] (0,-2)--(0,0);
 \draw (0,0)--(1,1);
 \draw (0,0)--(-1,1);
 \draw (3, -2)--(3,1);
 \draw (0,-2) node[anchor=east] {$\mathbf{1}$};
 \draw (3,-2) node[anchor=west] {${X}$};
\end{scope}
\begin{scope}[yshift=3cm]
 \draw (1,-2)--(-1,1);
 \draw [white, line width=3mm] (-1,-2)--(1,1);
 \draw (-1,-2)--(1,1);
 \draw (3, -2)--(3,1);
\end{scope}
\begin{scope}[yshift=6cm]
 \draw (3,-2)--(1,1);
 \draw [white, line width=3mm] (1,-2)--(3,1);
 \draw (1,-2)--(3,1);
 \draw (-1, -2)--(-1,1);
\end{scope}
\begin{scope}[yshift=9cm]
 \draw (-1,-2)--(1,1);
 \draw [white, line width=3mm] (1,-2)--(-1,1);
 \draw (1,-2)--(-1,1);
 \draw (3, -2)--(3,1);
\end{scope}
\begin{scope}[yshift=12cm]
 \draw (1,-2)--(3,1);
 \draw [white, line width=3mm] (3,-2)--(1,1);
 \draw (3,-2)--(1,1);
 \draw (-1, -2)--(-1,1);
\end{scope}
\begin{scope}[yshift=14cm]
 \draw (0,0)--(0,2);
 \draw (-1,-1)--(0,0);
 \draw (1,-1)--(0,0);
 \draw (3, -1)--(3,2);
 \draw (0,2) node[anchor=east] {${Y}$};
 \draw (3,2) node[anchor=west] {${X}$};
\end{scope}
 \end{scope}
\begin{scope}[xshift=8cm,yshift=7cm]
 \draw (0,0) node {$=~~\gamma$};
\end{scope}
\begin{scope}[xshift=12cm,yshift=4cm]
\begin{scope}[yshift=0cm]
 \draw [dashed] (0,-2)--(0,0);
 \draw (0,0)--(1,1);
 \draw (0,0)--(-1,1);
 \draw (3, -2)--(3,1);
 \draw (0,-2) node[anchor=east] {$\mathbf{1}$};
 \draw (3,-2) node[anchor=west] {${X}$};
\end{scope}
\begin{scope}[yshift=3cm]
 \draw (-1,-2)--(-1,2);
 \draw (1, -2)--(1,2);
 \draw (3, -2)--(3,2);
\end{scope}
\begin{scope}[yshift=6cm]
 \draw (0,0)--(0,2);
 \draw (-1,-1)--(0,0);
 \draw (1,-1)--(0,0);
 \draw (3, -1)--(3,2);
 \draw (0,2) node[anchor=east] {${Y}$};
 \draw (3,2) node[anchor=west] {${X}$};
\end{scope}
\end{scope}
\begin{scope}[xshift=19cm,yshift=7cm]
 \draw (0,0) node {$=~~0$};
\end{scope}
\end{tikzpicture}
\label{fig:graph is zero}
\caption{Self-dual case: a vanishing state}
\end{center}
\end{figure}

This contradiction implies that the image matrix of $b$ is not a scalar.  It follows that the images of the elementary braids $\sigma_1$ and $\sigma_2$ do not commute.

If $X$ is non-self-dual, then we will show that there is a braid in $B_4$ whose image is not the identity up to an overall scalar.

Consider the following braid $b'$ in the $4$-strand braid group $B_4$ (See Fig 4).

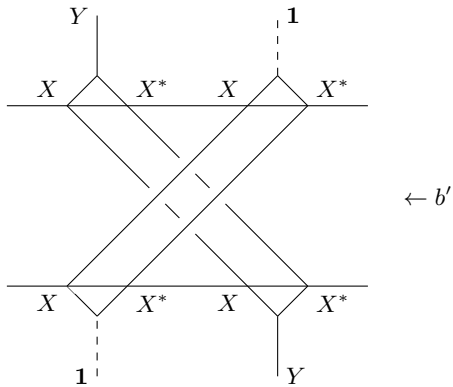
\begin{figure}[th!]
\begin{center}
\begin{tikzpicture}[scale=0.4]
  \begin{scope}[yshift=0cm]
    \draw [dashed](0,-2)--(0,0);
    \draw (0,0)--(1,1);
    \draw (0,0)--(-1,1);
    \draw (0,-2) node[anchor=east] {$\mathbf{1}$};
    \draw (-1,1) node[anchor=north east] {${X}$};
    \draw (1,1) node[anchor=north west] {${X}^*$};
  \end{scope}
  \begin{scope}[xshift=6cm]
    \draw  (0,-2)--(0,0);
    \draw (0,0)--(1,1);
    \draw (0,0)--(-1,1);
    \draw (0,-2) node[anchor=west] {${Y}$};
    \draw (-1,1) node[anchor=north east] {${X}$};
    \draw (1,1) node[anchor=north west] {${X}^*$};
  \end{scope}
  \begin{scope}[xshift=-1cm,yshift=1cm]
    \draw (8,0)--(2,6);
    \draw (6,0)--(0,6);
    \draw [white, line width=3mm] (0,0)--(6,6);
    \draw (0,0)--(6,6);
    \draw [white, line width=3mm] (2,0)--(8,6);
    \draw (2,0)--(8,6);
    \draw (-2, 0)--(10,0);
    \draw (-2, 6)--(10,6);
  \end{scope}
  \begin{scope}[yshift=8cm]
    \draw (0,0)--(0,2);
    \draw (-1,-1)--(0,0);
    \draw (1,-1)--(0,0);
    \draw (0,2) node[anchor=east] {${Y}$};
    \draw (-1,-1) node[anchor=south east] {${X}$};
    \draw (1,-1) node[anchor=south west] {${X}^*$};
  \end{scope}
  \begin{scope}[xshift=6cm, yshift=8cm]
    \draw [dashed] (0,0)--(0,2);
    \draw (-1,-1)--(0,0);
    \draw (1,-1)--(0,0);
    \draw (0,2) node[anchor=west] {$\mathbf{1}$};
    \draw (-1,-1) node[anchor=south east] {${X}$};
    \draw (1,-1) node[anchor=south west] {${X}^*$};
  \end{scope}
\begin{scope}[xshift=11cm,yshift=4cm]
 \draw (0,0) node {$\leftarrow{b'}$};
\end{scope}
\end{tikzpicture}

\label{fig:nsd case}
\caption{The non-self-dual case}
\end{center}
\end{figure}

Similarly to the argument above, on one hand we have the identity in Fig. 5, which shows the resulting state is non-zero.
\begin{figure}[th!]
\begin{center}
\begin{tikzpicture}[scale=0.27]
 \begin{scope}
 \begin{scope}[yshift=0cm]
    \draw [dashed](0,-2)--(0,0);
    \draw (0,0)--(1,1);
    \draw (0,0)--(-1,1);
    \draw (0,-2) node[anchor=east] {$\mathbf{1}$};
    \draw (-1,1) node[anchor=north east] {${X}$};
    \draw (1,1) node[anchor=north west] {${X}^*$};
  \end{scope}
  \begin{scope}[xshift=6cm]
    \draw  (0,-2)--(0,0);
    \draw (0,0)--(1,1);
    \draw (0,0)--(-1,1);
    \draw (0,-2) node[anchor=west] {${Y}$};
    \draw (-1,1) node[anchor=north east] {${X}$};
    \draw (1,1) node[anchor=north west] {${X}^*$};
  \end{scope}
  \begin{scope}[xshift=-1cm,yshift=1cm]
    \draw (8,0)--(2,6);
    \draw (6,0)--(0,6);
    \draw [white, line width=3mm] (0,0)--(6,6);
    \draw (0,0)--(6,6);
    \draw [white, line width=3mm] (2,0)--(8,6);
    \draw (2,0)--(8,6);
  \end{scope}
  \begin{scope}[yshift=8cm]
    \draw (0,0)--(0,2);
    \draw (-1,-1)--(0,0);
    \draw (1,-1)--(0,0);
    \draw (0,2) node[anchor=east] {${Y}$};
    \draw (-1,-1) node[anchor=south east] {${X}$};
    \draw (1,-1) node[anchor=south west] {${X}^*$};
  \end{scope}
  \begin{scope}[xshift=6cm, yshift=8cm]
    \draw [dashed] (0,0)--(0,2);
    \draw (-1,-1)--(0,0);
    \draw (1,-1)--(0,0);
    \draw (0,2) node[anchor=west] {$\mathbf{1}$};
    \draw (-1,-1) node[anchor=south east] {${X}$};
    \draw (1,-1) node[anchor=south west] {${X}^*$};
  \end{scope}
 \end{scope}
\begin{scope}[xshift=11cm,yshift=4cm]
 \draw (0,0) node {$=~~\alpha$};
\end{scope}
\begin{scope}[xshift=16cm, yshift=1cm]
  \begin{scope}[yshift=0cm]
   \draw [thick] (0,-1.5)--(0,0);
   \draw (0,0)--(1.5,1.5);
   \draw (0,0)--(-1.5,1.5);
   \draw (0,-1.5) node[anchor=east] {${Y}$};
   \draw (-1.5,1.5) node[anchor=east] {${X}$};
   \draw (1.5,1.5) node[anchor=west] {${X^*}$};
  \end{scope}
  \begin{scope}[yshift=3cm]
   \draw (-1.5,-1.5)--(-1.5,1.5);
   \draw (1.5, -1.5)--(1.5,1.5);
  \end{scope}
 \begin{scope}[yshift=6cm]
  \draw (0,0)--(0,1.5);
  \draw (-1.5,-1.5)--(0,0);
  \draw (1.5,-1.5)--(0,0);
  \draw (0,1.5) node[anchor=east] {${Y}$};
\end{scope}
\end{scope}
\begin{scope}[xshift=22cm,yshift=4cm]
 \draw (0,0) node {$\neq~~0$};
\end{scope}
\end{tikzpicture}
\label{fig:nsd case not zero}
\caption{Non-self-dual: the overstrands result in a non-zero scalar.  The right-hand side is a non-zero state.}
\end{center}
\end{figure}
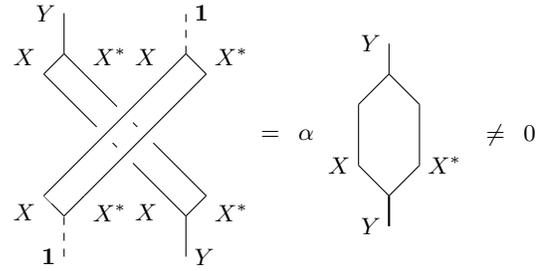
On the other hand, if the image of the middle braid $b'$ is the same as the identity up to an overall phase, then Fig. 6 implies that the resulting state would be $0$.

\begin{figure}[th!]
\begin{center}
\begin{tikzpicture}[scale=0.28]
 \begin{scope}
  \begin{scope}[yshift=0cm]
    \draw [dashed](0,-2)--(0,0);
    \draw (0,0)--(1,1);
    \draw (0,0)--(-1,1);
    \draw (0,-2) node[anchor=east] {$\mathbf{1}$};
    \draw (-1,1) node[anchor=north east] {${X}$};
    \draw (1,1) node[anchor=north west] {${X}^*$};
  \end{scope}
  \begin{scope}[xshift=6cm]
    \draw  (0,-2)--(0,0);
    \draw (0,0)--(1,1);
    \draw (0,0)--(-1,1);
    \draw (0,-2) node[anchor=west] {${Y}$};
    \draw (-1,1) node[anchor=north east] {${X}$};
    \draw (1,1) node[anchor=north west] {${X}^*$};
  \end{scope}
  \begin{scope}[xshift=-1cm,yshift=1cm]
    \draw (8,0)--(2,6);
    \draw (6,0)--(0,6);
    \draw [white, line width=3mm] (0,0)--(6,6);
    \draw (0,0)--(6,6);
    \draw [white, line width=3mm] (2,0)--(8,6);
    \draw (2,0)--(8,6);
  \end{scope}
  \begin{scope}[yshift=8cm]
    \draw (0,0)--(0,2);
    \draw (-1,-1)--(0,0);
    \draw (1,-1)--(0,0);
    \draw (0,2) node[anchor=east] {${Y}$};
    \draw (-1,-1) node[anchor=south east] {${X}$};
    \draw (1,-1) node[anchor=south west] {${X}^*$};
  \end{scope}
  \begin{scope}[xshift=6cm, yshift=8cm]
    \draw [dashed] (0,0)--(0,2);
    \draw (-1,-1)--(0,0);
    \draw (1,-1)--(0,0);
    \draw (0,2) node[anchor=west] {$\mathbf{1}$};
    \draw (-1,-1) node[anchor=south east] {${X}$};
    \draw (1,-1) node[anchor=south west] {${X}^*$};
  \end{scope}
 \end{scope}
\begin{scope}[xshift=10cm,yshift=4cm]
 \draw (0,0) node {$=~~\gamma$};
\end{scope}
\begin{scope}[xshift=15cm, yshift=1cm]
  \begin{scope}[yshift=0cm]
   \draw [thick, dashed] (0,-1.5)--(0,0);
   \draw (0,0)--(1.5,1.5);
   \draw (0,0)--(-1.5,1.5);
   \draw (0,-1.5) node[anchor=east] {$\mathbf{1}$};
   \draw (-1.5,1.5) node[anchor=east] {${X}$};
   \draw (1.5,1.5) node[anchor=west] {${X^*}$};
  \end{scope}
  \begin{scope}[yshift=3cm]
   \draw (-1.5,-1.5)--(-1.5,1.5);
   \draw (1.5, -1.5)--(1.5,1.5);
  \end{scope}
 \begin{scope}[yshift=6cm]
  \draw (0,0)--(0,1.5);
  \draw (-1.5,-1.5)--(0,0);
  \draw (1.5,-1.5)--(0,0);
  \draw (0,1.5) node[anchor=east] {${Y}$};
\end{scope}
\end{scope}
\begin{scope}[xshift=22cm, yshift=1cm]
  \begin{scope}[yshift=0cm]
   \draw [thick] (0,-1.5)--(0,0);
   \draw (0,0)--(1.5,1.5);
   \draw (0,0)--(-1.5,1.5);
   \draw (0,-1.5) node[anchor=east] {${Y}$};
   \draw (-1.5,1.5) node[anchor=east] {${X}$};
   \draw (1.5,1.5) node[anchor=west] {${X^*}$};
  \end{scope}
  \begin{scope}[yshift=3cm]
   \draw (-1.5,-1.5)--(-1.5,1.5);
   \draw (1.5, -1.5)--(1.5,1.5);
  \end{scope}
 \begin{scope}[yshift=6cm]
  \draw [thick, dashed](0,0)--(0,1.5);
  \draw (-1.5,-1.5)--(0,0);
  \draw (1.5,-1.5)--(0,0);
  \draw (0,1.5) node[anchor=east] {$\mathbf{1}$};
\end{scope}
\end{scope}
\begin{scope}[xshift=27cm,yshift=4cm]
 \draw (0,0) node {$=~~0$};
\end{scope}
\end{tikzpicture}
\label{fig:nsd case zero}
\caption{Non-self-dual: replacing $b'$ by the identity braid results in a zero state.}
\end{center}
\end{figure}
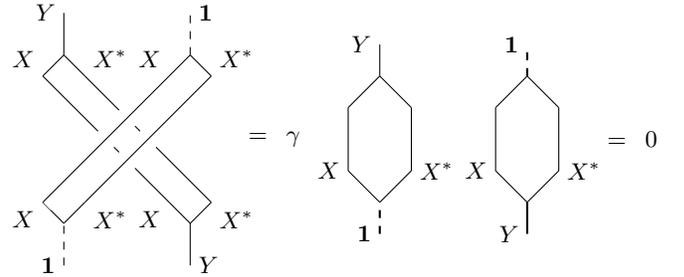
The contradiction implies that the image of the braid $b'$ is not a scalar.

\section{Conclusions}

In this letter, we find that degeneracy is more fundamental than non-abelian statistics.  One consequence is that experimental confirmation of non-abelian fusion rules implies non-abelian braiding statistics if anyon systems are modeled by TQFTs or unitary modular categories up to overall phases.

\acknowledgments

Rowell is partially supported by NSF grant DMS-1108725, and Wang by NSF grant DMS-1108736.  We thank Parsa Bonderson, Mike Freedman, and Xiao-Gang Wen for their interests and helpful comments.  In particular, Bonderson pointed it out that our result holds for premodular category without any change.  Wen has expressed the same point of view that degeneracy is more fundamental than non-abelian braiding \footnote{Answer to how non-abelian anyons arise in solid state systems at physics stackexchange.}.

\bibliography{NAstatistics}

\end{document}